\documentclass[pdflatex,sn-mathphys-num]{sn-jnl}

\usepackage{graphicx}%
\usepackage{multirow}%
\usepackage{amsmath,amssymb,amsfonts}%
\usepackage{amsthm}%
\usepackage{mathrsfs}%
\usepackage[title]{appendix}%
\usepackage{xcolor}%
\usepackage{textcomp}%
\usepackage{manyfoot}%
\usepackage{booktabs}%
\usepackage{algorithm}%
\usepackage{algorithmicx}%
\usepackage{algpseudocode}%
\usepackage{listings}%
\usepackage{graphics}%
\usepackage{tabularx}%
\usepackage{float}%
\usepackage{subcaption}%
\usepackage{threeparttable}%
\usepackage{pdflscape}%

\usepackage[colorinlistoftodos,textsize=tiny,textwidth=14mm]{todonotes} 
\usepackage{soul}%
\usepackage{xspace}%
\usepackage{adjustbox}%

\theoremstyle{thmstyleone}%
\theoremstyle{thmstyletwo}%

\theoremstyle{thmstylethree}%

\begin{document}

\title[Article Title]{Shifting Social Dispositions, Stable Prosocial Traits: A Global Age-Period-Cohort Analysis of Human Personality}


\author*[1,2,3]{\fnm{Paul X.} \sur{McCarthy}}\email{paul@leagueofscholars.com}
\equalcont{These authors contributed equally to this work.}

\author[2,3]{\fnm{Xian} \sur{Gong}}\email{xian@leagueofscholars.com}
\equalcont{These authors contributed equally to this work.}

\author[4]{\fnm{John A.} \sur{Johnson}}\email{j5j@psu.edu}

\author[2]{\fnm{Marian-Andrei} \sur{Rizoiu}}\email{Marian-Andrei.Rizoiu@uts.edu.au}

\author[5]{\fnm{Margaret L.} \sur{Kern}}\email{Peggy.Kern@unimelb.edu.au}

\author[6]{\fnm{Jean M.} \sur{Twenge}}\email{jtwenge@sdsu.edu}

\affil*[1]{\orgdiv{Computer Science and Engineering}, \orgname{UNSW Sydney}, \orgaddress{\city{Sydney}, \postcode{2052}, \state{NSW}, \country{Australia}}}

\affil[2]{\orgname{League of Scholars}, \orgaddress{\city{Sydney}, \postcode{1225}, \state{NSW}, \country{Australia}}}

\affil[3]{\orgdiv{Faculty of Engineering and Information Technology}, \orgname{University of Technology Sydney}, \orgaddress{\city{Sydney}, \postcode{2007}, \state{NSW}, \country{Australia}}}

\affil[4]{\orgdiv{Department of Psychology}, \orgname{Pennsylvania State University}, \orgaddress{\city{University Park}, \postcode{16802}, \state{PA}, \country{United States}}}

\affil[5]{\orgdiv{Faculty of Education}, \orgname{University of Melbourne}, \orgaddress{\city{Melbourne}, \postcode{3010}, \state{Victoria}, \country{Australia}}}

\affil[6]{\orgdiv{Department of Psychology}, \orgname{San Diego State University}, \orgaddress{\city{San Diego}, \postcode{92182}, \state{CA}, \country{United States}}}


\abstract{Generational stereotypes are widespread, but they often rely on anecdotes, and it remains challenging to disentangle true birth-cohort differences from the universal effects of ageing and historical periods. Using a statistical approach that separates effects of age, calendar time, and cohort, we analyzed Big Five personality data (five broad dimensions of personality) from N=773,714 individuals assessed across 30 years. Traits that shape social interaction diverged between generations, whereas a prosocial core comprising morality, discipline, and emotional awareness was stable across cohorts. Generation Z (born between 1995 and 2012) showed lower excitement seeking and gregariousness alongside higher self-consciousness and anxiety. These findings suggest that cohort change is selective rather than global: the dispositions through which people engage with the social world may be adapting to contemporary cultural conditions, while core prosocial tendencies remain stable across generations.}

\keywords{Age-Period-Cohort Analysis, Generational Cohorts, Big Five Personality, Intergenerational Shifts}



\maketitle

\section{Introduction}\label{sec1}
From claims that Millennials are unusually entitled to portrayals of Generation Z as intrinsically anxious, popular narratives have long attributed distinctive personality profiles to different generations. Recent cohorts exhibit meaningful psychological and behavioural shifts shaped by technological and societal change \cite{twenge2023generations}. However, analysing personality differences across generations poses substantial methodological challenges, most notably the difficulty of disentangling age effects, period (historical time) effects, and cohort effects, which are inherently confounded in observational data \cite{costa1982attri, buss1991evol, mischel2008intro, roemer2026beyond}. Consequently, an ongoing debate in psychology and sociology questions whether measurable, large-scale differences between generations actually exist once age and period effects are properly accounted for. Although survey evidence supports cohort-based transformations \cite{twenge2017igen}, other empirical studies suggest that broad societal changes exert only a minor influence over the distribution of personality traits once age and contextual factors are considered \cite{rudolph2021generations, costanza2017review}. 

Personality traits exhibit normative developmental change across the life course \cite{soto2011age}. For example, as people age, emotional stability, agreeableness, and conscientiousness are likely to increase. This phenomenon creates the risk that age-related maturation is misinterpreted as generational change \cite{baltes1968longitudinal, ion2022secular}. Cultural specificity and historical contingency further complicate comparisons, as generational labels and their associated meanings vary across societies and time periods \cite{lyons2014generation, lukianoff2018coddling}. Further, many studies rely on cross-sectional designs or arbitrary generational boundaries, limiting causal inference and exaggerating apparent cohort differences \cite{gentile2015problem, costanza2017review, strickhouser2017personality}, as cohort effects cannot be separated from age-related development or historical period influences \cite{baltes1968longitudinal, Schmidt2002TheRO, stelling2023applicants}.

The fundamental challenge in generational research is the identification problem: how to distinguish amongst age, period, and cohort effects (Age = Period - Cohort). Previous studies could not reliably break this collinearity, often conflating the biological maturation process with concurrent historical events. The use of broad and culturally contingent generational labels further introduces conceptual ambiguity, as birth-year cut-offs are often arbitrary and inconsistently defined across studies and societies \cite{parry2011generation, rudolph2020cross}. In addition, effect sizes associated with generational differences in personality traits are typically small and sensitive to model specification, raising concerns about statistical over interpretation and publication bias \cite{kowske2010millenniais, funder2019eval, maehler2024cultural, twenge2010birth}. Finally, most large-scale datasets are drawn from Western populations, limiting the generalizability of findings and obscuring potential cross-cultural variation in personality development and cohort dynamics \cite{hayward2015aging, tsai2025postmaterialism, mockaitis2025between}. 

Overcoming these challenges requires massive, multi-decade sampling, from participants of different ages over many years and integrative frameworks that explicitly model developmental, historical, and cultural processes \cite{akaliyski2025develop}. Importantly, the recent accumulation of global survey data spanning decades presents a historically unique opportunity to isolate these variables. Our study draws on a large-scale, global personality dataset collected through an online IPIP-NEO survey. The instrument implements the International Personality Item Pool (IPIP) representation of the Five-Factor Model (five higher order personality factors plus 30 lower order facets), which comprises 300 questions with 10 items per facet. Additionally, the survey records basic demographic data, including respondents’ age, gender, and time taken to complete the questionnaire. The dataset spans three decades of data (1993–2023) and includes responses from more than 700,000 individuals worldwide.

Beyond the methodological challenge of separating age, period, and cohort effects, there is an important substantive question: are all parts of personality equally historically variable? One possibility is that cohort change is selective rather than uniform. Traits related to cooperation, emotional awareness, and self-regulation may remain comparatively stable across generations, whereas traits governing sociability, stimulation-seeking, and social vigilance may be more responsive to changing developmental environments. The present study tests this possibility using a large global sample and models that explicitly separate maturation, historical period, and birth-cohort effects.

For analytical purposes, respondents are grouped into generational cohorts based on birth year, including the Greatest Generation (born 1901–1924), The Silent Generation (1925-1945), Baby Boomers (1946-1964), Generation X (1965-1979), Millennials (1980-1994), and Generation Z (1995–2012), with cohort sample sizes varying substantially (see \ref{tab:tab1}). While sample sizes in the oldest generations are comparatively small, there are robust and large samples of the next four generations. This breadth of coverage across age groups and birth cohorts enables fine-grained analysis of personality traits and facets across the life course, while also enabling careful consideration of cohort imbalance and age-related confounding in subsequent analyses.

\begin{table}
  \caption{\textbf{Sample size and birth years of generational cohorts.} \textmd{Participant distribution from the Global Big5 IPIP300 Personality Survey (1993–2023) categorized by standard generation definitions. The table outlines the birth years, sample size (total N = 773,714), and the biological age range of respondents at the time of survey completion for each of the six cohorts analysed.}}
  \label{tab:tab1}
  \centering
  \begin{tabular}{lccc}
    \toprule[1pt]\midrule[0.3pt]
    \centering\textbf{Generations} &
    \textbf{Birth Years} &
    \textbf{Sample Size} &
    \textbf{Age Range at Survey} \\
    \midrule[0.5pt]
    The Greatest Generation & 1901-1924 & 199 & 77-99 \\
    The Silent Generation & 1925-1945 & 2,628 & 56-97 \\
    Baby Boomers & 1946-1964 & 43,107 & 37-76 \\
    Generation X & 1965-1979 & 125,790 & 22-57 \\
    Millennials & 1980-1994 & 349,217 & 10-42 \\
    Generation Z & 1995-2012 & 252,773 & 10-27 \\
  \midrule[0.3pt]\bottomrule[1pt]
\end{tabular}
\end{table}

\section{Results}\label{sec2}
To separate age (maturation), period (year), and cohort (generational) effects, we analyzed a global dataset of 773,714 participants using Hierarchical Age-Period-Cohort (H-APC) \cite{yang2008age, reither2015clarifying, BellAndrew2018Tham} modelling, which is a multilevel approach that treats period and cohort as higher-level (cross-classified) random effects, enabling researchers to estimate age, period, and cohort effects simultaneously while partially addressing the identification problem. Our analysis reveals a complex picture: while generations exhibit marked divergence on many personality dimensions, on a small number of others, they are virtually identical.

\subsubsection*{The Stable Trio: An Unchanging Prosocial Core}
\begin{figure}
	\centering
	\includegraphics[width=\textwidth]{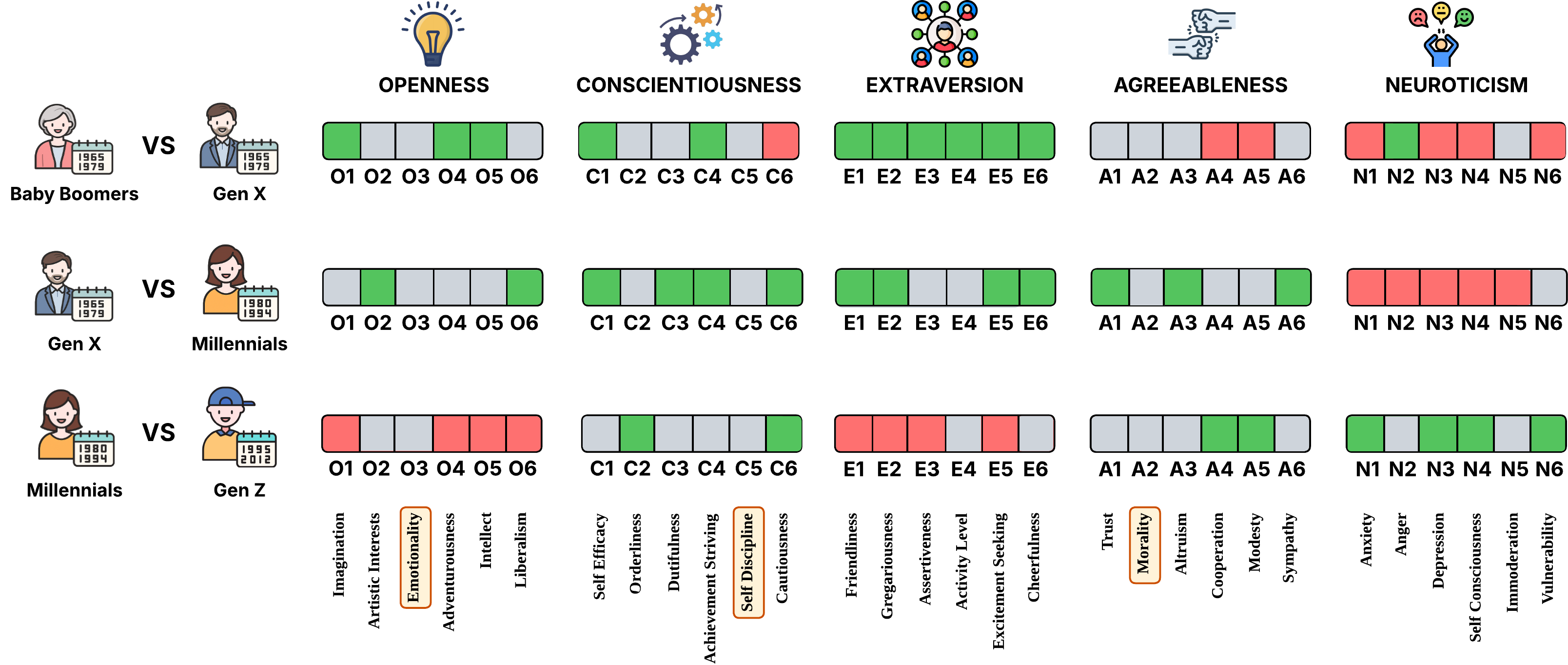}
	\caption{\textbf{Intergenerational commonalities and differences across the Big Five personality domains.} \textmd{Heatmaps illustrate the magnitude and direction of personality trait differences across consecutive generation pairs, adjusted for age and period effects. Statistically significant (p $\leq$ 0.05) notable increases from the older generation to the younger generation are shown in green, notable decreases in red, and negligible or marginal changes in grey. The analysis reveals a resilient "Stable Trio" (O3, C5, A2), which resists generational shifts, contrasting sharply with the volatility evident in Extraversion, Neuroticism, and Cautiousness (C6).}}
	\label{fig:fig1} 
\end{figure}

As illustrated in Figure~\ref{fig:fig1}, which illustrates heat-maps adjusted for age and period effects, a specific cluster of personality facets, which we term the ``Stable Trio", emerges as being consistent across the last four generations. This cluster consists of the personality facets of Emotionality (O3), Self-Discipline (C5), and Morality (A2). The stability of these traits supports the evolutionary perspective that the Stable Trio act as the ``biological hardware" of human nature. From this perspective, fairness (from Morality), impulse control (from Self-Discipline), and the ability to regulate emotions (from Emotionality) are important for maintaining cooperative alliances amongst individuals. Fairness reflects the consistent pursuit of justice, even as social structures fluctuate. Self-Discipline reflects the internal determination necessary to overcome hardship. Emotionality reflects the ability to be aware of and open to emotion, suggesting that the ``open wound" of human sensitivity is a timeless feature of our species' emotional architecture \cite{lane1990levels}.

\subsubsection*{Intergenerational Volatility and Permeable Traits}
In contrast to the Stable Trio, the H-APC analysis demonstrates that personality development occurs through compound trends. Traits that impact how individuals interface with groups and process environmental stimuli are permeable and variable. These are the ``user interface" settings of the personality, adjusting dynamically to the cultural and historical ``operating environment" rather than to a biological baseline.

\begin{figure}
     \centering
     \begin{subfigure}[a]{\textwidth}
         \centering
         \includegraphics[width=\textwidth]{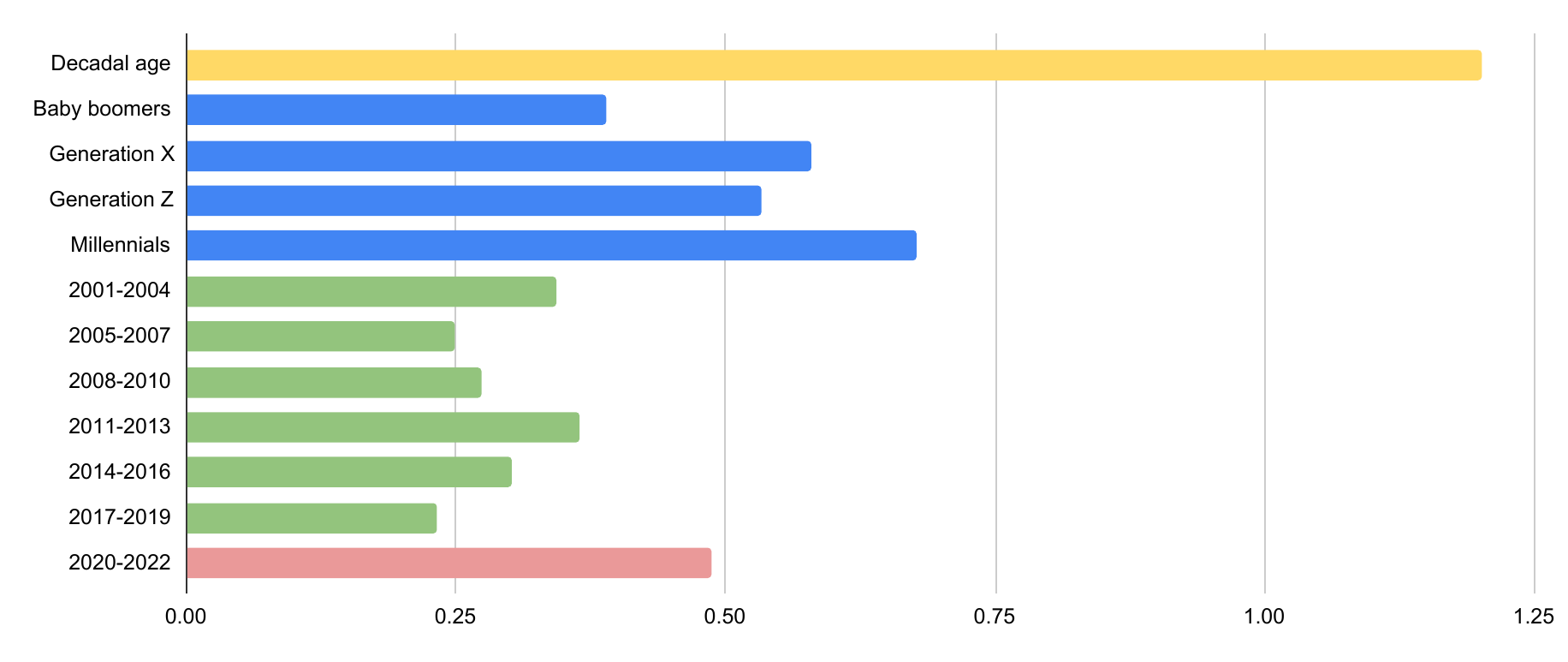}
         \caption{}
         \label{fig:3a}
     \end{subfigure}
     \vfill
     \begin{subfigure}[b]{\textwidth}
         \centering
         \includegraphics[width=0.9\textwidth]{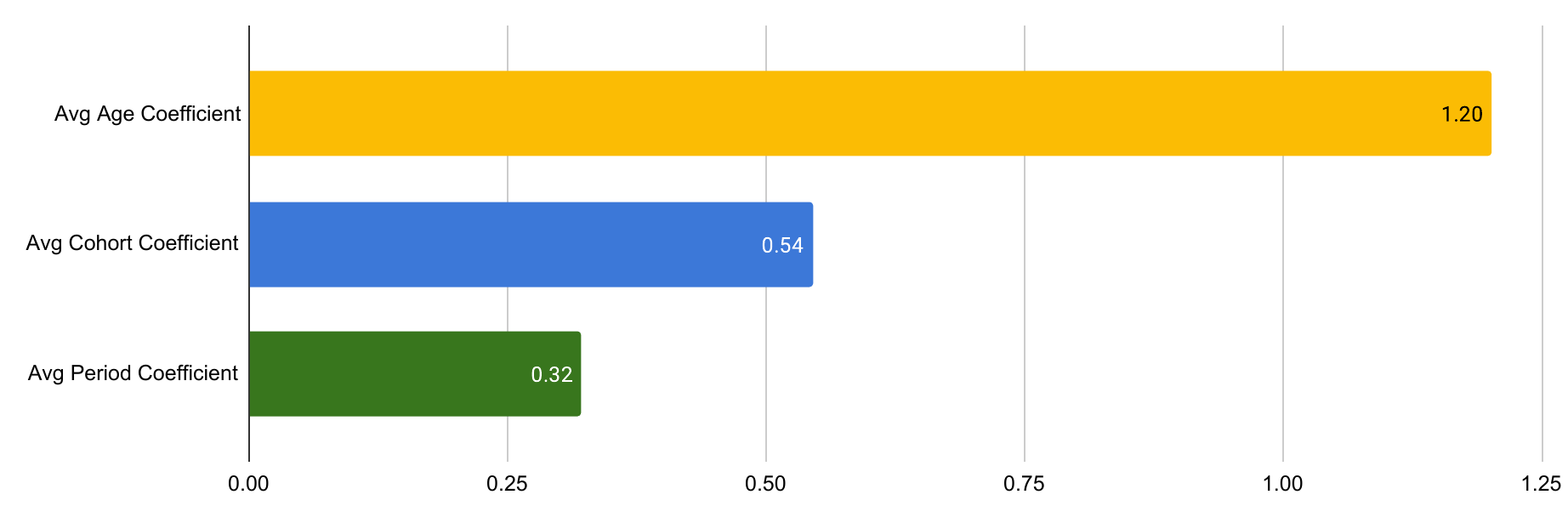}
         \caption{}
         \label{fig:3b}
     \end{subfigure}
        \caption{\textbf{Comparative influence of maturation, cohort, and period on personality traits.} (A) Absolute average standardized coefficients from the H-APC model compare key analytic dimensions. Decadal age represents the biological age effect scaled by ten years to allow for direct magnitude comparison. (B) Aggre-gated summary confirms maturation (Avg Age Coefficient) is the primary driver of personality variance, followed by generational cohort and then historical period.}
        \label{fig:fig2}
\end{figure}

For instance, hierarchical age-period-cohort data reveal a quantifiable intergenerational decline in Excitement Seeking (E5) and Gregariousness (E2), whereas, Self-Consciousness (N4) and Anxiety (N1) increase over the generations. Analysis of survey years 2001-2022 confirms that historical period effects exert a comparatively negligible influence on these traits, justifying the analytical focus on age and birth cohort as the primary drivers of variance. As detailed in Figure~\ref{fig:fig2}, the aggregated summary confirms that maturation (ageing) is the primary driver of personality variance, followed by the generational cohort. The average cohort coefficient (0.54) significantly outweighs the aggregate time period coefficient (0.32), confirming that birth cohort is the dominant environmental driver of these trait shifts.

\begin{figure}
	\centering
	\includegraphics[width=\textwidth]{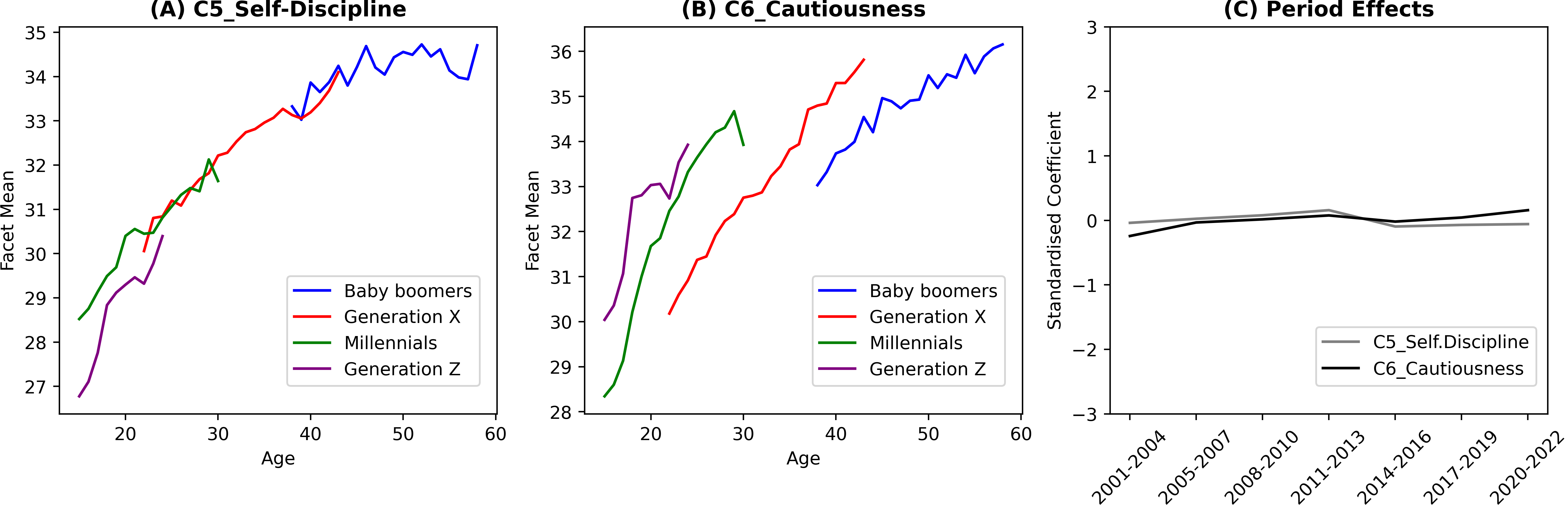}
	\caption{\textbf{Direct age-paired sampling isolates maturational and generational effects.} \textmd{(A) Direct age-paired comparisons of individuals (n = 670,008) from consecutive generations at identical ages for Self-Discipline (C5) illustrate a pure maturation effect, with generational trajectories overlapping seamlessly to demonstrate strict intergenerational stability. (B) In contrast, age-to-age comparisons for Cautiousness (C6) reveal a compound trend where all cohorts exhibit age-related development, yet successive younger generations show significantly elevated baseline cautiousness at identical ages, exposing a distinct generational step-change. (C) Analysis of survey years 2001–2022 confirms that historical period effects exert a comparatively negligible influence on these traits, justifying the analytical focus on age and birth cohort as the primary drivers of variance.}}
	\label{fig:fig3} 
\end{figure}

Further, as illustrated in Figure~\ref{fig:fig3}, the personality facet of cautiousness (C6) shows a clear maturation effect, and a distinct cohort effect: successive younger generations exhibit significantly elevated baseline Cautiousness levels compared to older cohorts at equivalent ages. For example, Generation Z individuals at age 25 display the Cautiousness levels typical of a 35-year-old Generation X individual or a 40-year-old Baby Boomer. In contrast, Self-Discipline (C5) illustrates a pure maturation effect, with generational trajectories overlapping seamlessly to demonstrate strict intergenerational stability.

\section{Discussion}\label{sec3}
The divergent trajectories of personality development observed in this large-scale, global dataset provides a framework for understanding human stability and adaptation. Although some traits are stable over the ages (evidenced here as our Stable Trio), our social mode appears to be responsive to technological and cultural changes in the environment. This suggests that while our fundamental prosocial traits remain immutable, the behavioural mechanisms through which individuals engage with society are subject to profound structural transformation.

The behavioural shifts observed in Generation Z provide empirical support for Twenge's slow life strategy hypothesis and Haidt's framework of safetyism, as well as previous time-lag studies finding increases in anxiety and depression among adolescents since 2012. Previous cohorts sought novelty through physical risk-taking, whereas younger generations appear to be adapting to a substantially altered developmental environment, driven by rapid technological and sociocultural shifts detailed below. The timing of these pronounced shifts aligns with the structural transformation of childhood and adolescence. While multiple compounding sociocultural factors likely contribute to this transition, global smartphone penetration crossing the 50 percent threshold in the early 2010s serves as a highly probable catalyst. This structural transition maps directly onto the accelerated decline in Extraversion and the growth in Neuroticism observed in our Generation Z sample. The demands of maintaining curated digital identities likely generate performative anxiety, shifting social interaction from a source of emotional closeness to a continuous environmental stressor\cite{twenge2017igen,lukianoff2018coddling,twenge2023generations}.

However, the etiology of these cohort effects is undoubtedly complex. Further research is required to fully disentangle the specific impact of ubiquitous digital technologies from other concurrent macro-societal shifts that may have primed early childhood environments for heightened risk aversion, including increases in natural disasters, political instability, dysfunctional family environments, and psychological pressures from the Covid-19 pandemic.

Importantly, the data indicate that modern technology and broader cultural shifts have not fundamentally altered the Stable Prosocial Core of basic human values. They have instead driven a pronounced rewiring of our social dispositions, effectively substituting dopamine-driven risk and excitement with cortisol-driven anxiety and an elevated baseline of cautiousness.

This investigation was catalysed by our prior research into the determinants of high-growth ventures \cite{mccarthy2023impact}, where we established that successful founders possess a distinct phenotypic signature characterized by high Openness, specifically the Adventurousness facet, and high Extraversion. The generational divergence we observe in these same traits presents a critical economic signal. With Generation Z lower in Adventurousness and Excitement Seeking than previous generations, we face a potential demographic bottleneck for future entrepreneurship. The psychological fuel for high-risk, high-reward venture creation is becoming scarcer. This suggests the innovation economy may be shifting from a landscape of reckless experimentation to one of cautious optimization.

These findings also may contribute to discourse on ``grit", defined as passion and perseverance for long-term goals~\cite{duckworth2016grit}. Self-Discipline, which was part of the Stable Trio, reflects the perseverance part of grit, suggesting that grit is partially a fixed part of our biology. However, other traits such as Anxiety (N1) and Self-Consciousness (N4), may obstruct passion and stand in the way of gritty behaviours. This diverts cognitive resources toward emotional regulation and risk management rather than external achievement. This would suggest that the capacity for grit remains intact, yet it is currently overwhelmed by the noise of the modern social environment.

Finally, these generational shifts carry significant implications for the broader economy, particularly regarding the necessary diversity of personality types within a population~\cite{mccarthy2025economics}. National prosperity is driven not just by institutions but by the diversity of personality types within a population. Complex economies require a functional heterogeneity: a mix of ``Pathfinders" to innovate, ``Uniters" to stabilize, and ``Strategists'' to optimize. The generational rewiring we document here poses a specific macroeconomic risk. If the population mean shifts too heavily toward high Neuroticism and low Extraversion, we risk eroding the optimal diversity required for economic dynamism. By shrinking the pool of risk-tolerant ``Pathfinder'' personalities, we may inadvertently be designing an economy that is stable but stagnant, lacking the psychological variance necessary to drive GDP growth and technological adaptation.

\section{Methods}\label{sec4}

\subsection{Dataset and Participant Selection}
We analyzed data from the Global Big5 IPIP NEO 300 Personality Survey~\cite{johnson2014measuring}. This dataset comprises an aggregated pool of 773,714 valid responses collected continuously between 1993 and 2023, from participants between the ages of 10 and 99. Participants represent over 100 countries, making it a comprehensive dataset. The instrument utilizes a 300-item questionnaire to measure 30 distinct personality facets on a five-point Likert scale across five personality factors (Openness, Conscientiousness, Extraversion, Agreeableness, Neuroticism). Participants were categorized into standard birth cohorts: The Greatest Generation (1901 to 1924; n=199), the Silent Generation (1925 to 1945; n=2,628), Baby Boomers (1946 to 1964; n=43,107), Generation X (1965 to 1979; n=125,790), Millennials (1980 to 1994; n=349,217), and Generation Z (1995 to 2012; n=252,773).

\subsection{Data Cleaning and Quality Control}
Because the dataset is a collation of smaller subsets over three decades, the raw data required stringent quality control. We removed duplicate submissions by cross-referencing chronological timestamps and matching participant identifiers. To eliminate careless responding, protocols demonstrating excessive consecutive use of a single response category (straight-lining) were excluded. Protocol internal consistency and consistent responding to psychometric antonyms were assessed to validate the reliability of each submission. Relevant items were reverse-scored prior to handling missing data. Surveys with excessive missing data were discarded. For surveys with isolated omissions, values were imputed to the scale midpoint.

\subsection{Hierarchical Age-Period-Cohort (H-APC) Modelling}
We applied H-APC modelling across all 30 facets to resolve the linear dependence among age, period, and cohort. This framework formally disentangles biological maturation from historical survey context and generational cohort membership. We extracted standardized coefficients and significance values (p <= 0.05) to quantify the magnitude and direction of trait shifts between consecutive generations. Standard thresholds for effect sizes based on Cohen's d were used to classify shifts as notable increases, notable decreases, or marginal trends. This classification generated the primary division between the stable prosocial core and the highly volatile trait domains.

\subsection{Age-Paired Sampling and Variance Partitioning}
To validate the modelled trajectories, we utilized a direct age-paired sampling technique. The three-decade span allowed the extraction of overlapping age ranges for adjacent cohorts. Individuals from successive generations were compared at identical biological ages. For example, Baby Boomers were matched to Generation X participants at the exact same age. Holding age constant while varying the survey window isolates pure maturation from compound generational step-changes. 

Finally, we aggregated the absolute average standardized coefficients from the H-APC model to assess the relative influence of the three variables on overall personality variance. The biological age effect was scaled by ten years to allow direct magnitude comparisons. Survey period effects were modelled in three-year windows from 2001 to 2022. This permitted a definitive quantification of the comparative weights of biological maturation, generational membership, and historical era.

\section{Data availability}\label{sec5}
All data needed to evaluate the conclusions in the paper are present in the paper and/or the Supplementary Materials. The raw data from the Global Big5 IPIP-NEO 300 Personality Survey and the code used to run the Hierarchical Age-Period-Cohort (H-APC) modelling will be deposited in a publicly available repository upon publication of this manuscript.

\section{Code availability}\label{sec6}
All code scripts for the paper are available on GitHub.

\bibliography{sn-bibliography}

\section{Acknowledgments}\label{sec7}
We thank Fabian Braesemann (University of Oxford) for posing the initial question regarding intergenerational variances in entrepreneurial personality traits, which catalysed this project, and for his generous comments and review of the manuscript. We are grateful to Steven Pinker for his kind introduction to Jean M. Twenge, which helped bring this collaboration together. We also thank Mark McCrindle for his valuable insights and encouragement during the early stages of this research.

\section{Author contributions}\label{sec8}
All authors had substantial input into the project, and all authors read and approved the final manuscript. Individual contributions according to the CRediT taxonomy are as follows:
\begin{itemize}
  \item Conceptualization: PXM, XG
  \item Methodology: PXM, XG, MAR
  \item Formal Analysis: PXM, XG
  \item Investigation: PXM, XG
  \item Data Curation: JAJ
  \item Validation: JMT, MLK
  \item Visualization: PXM, XG
  \item Writing – Original Draft: PXM, XG
  \item Writing – Review \& Editing: PXM, XG, JAJ, MAR, MLK, JMT
\end{itemize}

\section{Competing interests}\label{sec10}
There are no competing interests to declare.

\section{Additional information}\label{sec11}
Supplementary information is available for this paper at the link.

\begin{appendices}

\section{}\label{secA1}

To provide complete transparency of the statistical modelling and allow for precise cross-cohort comparisons, this section details the full output of the Hierarchical Age-Period-Cohort (H-APC) analysis and standardizes the most significant intergenerational shifts using Cohen's d effect sizes.

\subsubsection*{Visualizing Intergenerational Volatility}
While the H-APC model coefficients confirm the primary drivers of personality variance, calculating standardized effect sizes clarifies the exact magnitude and direction of trait shifts between consecutive cohorts. Figures S1 through S3 highlight the ten facets exhibiting the largest positive and negative deviations across three distinct generational transitions.

\begin{figure} 
	\centering
	\includegraphics[width=\textwidth]{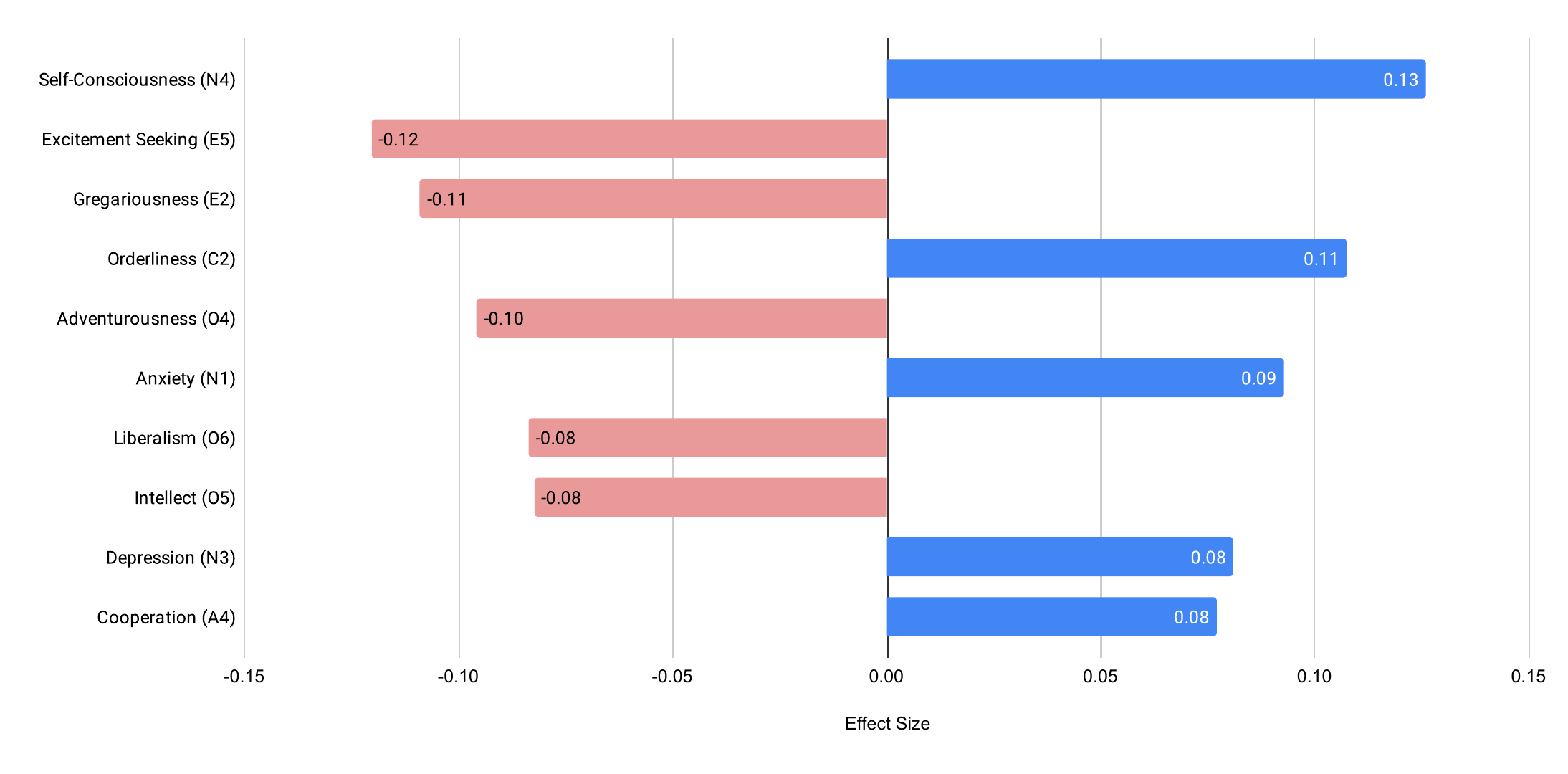} 

	\caption{\textbf{Biggest Intergenerational Changes: Generation Z compared to Millennials.} Standardized effect sizes (Cohen’s d) illustrating the largest differences between Millennials and Generation Z. Gen Z scores higher than Millennials on the traits shown with blue bars and lower on those shown with pink bars The data reveal a structural shift toward inward focus and risk aversion. Generation Z demonstrates substantial increases in Neuroticism facets, led by Self-Consciousness (d = 0.13) and Anxiety (d = 0.09). Concurrently, the largest declines appear in Extraversion and Openness facets, specifically Excitement Seeking (d = -0.12), Gregariousness (d = -0.11), and Adventurousness (d = -0.10).}
	\label{fig:sup_fig1} 
\end{figure}

\begin{figure} 
	\centering
	\includegraphics[width=\textwidth]{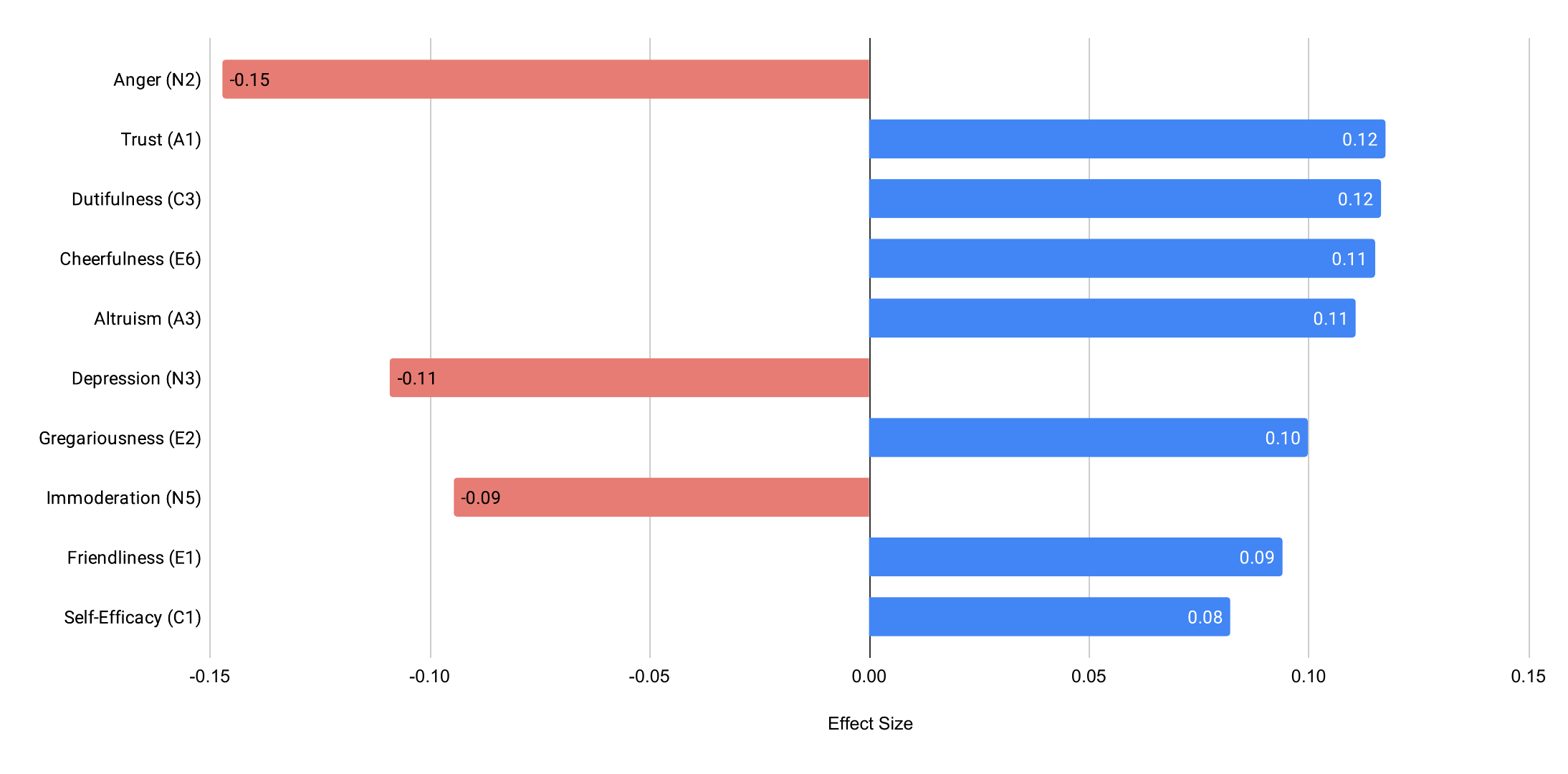} 

	\caption{\textbf{Biggest Intergenerational Changes: Millennials compared to Generation X.} Standardized effect sizes (Cohen’s d) illustrating the largest magnitude trait shifts between Generation X and Millennials. Millennials score higher than Gen X on the traits shown with blue bars and lower on those shown with pink bars. In contrast to the subsequent Generation Z transition, the shift from Gen X to Millennials is characterized by enhanced emotional stability and prosocial orientation. Anger exhibited the most pronounced absolute decline (d = -0.15), alongside measurable decreases in Depression (d = -0.11) and Immoderation (d = -0.09). Concurrently, Millennials registered elevated levels of Trust (d = 0.12), Dutifulness (d = 0.12), Cheerfulness (d = 0.11), and Altruism (d = 0.11) relative to Generation X.}
	\label{fig:sup_fig2} 
\end{figure}

\begin{figure} 
	\centering
	\includegraphics[width=\textwidth]{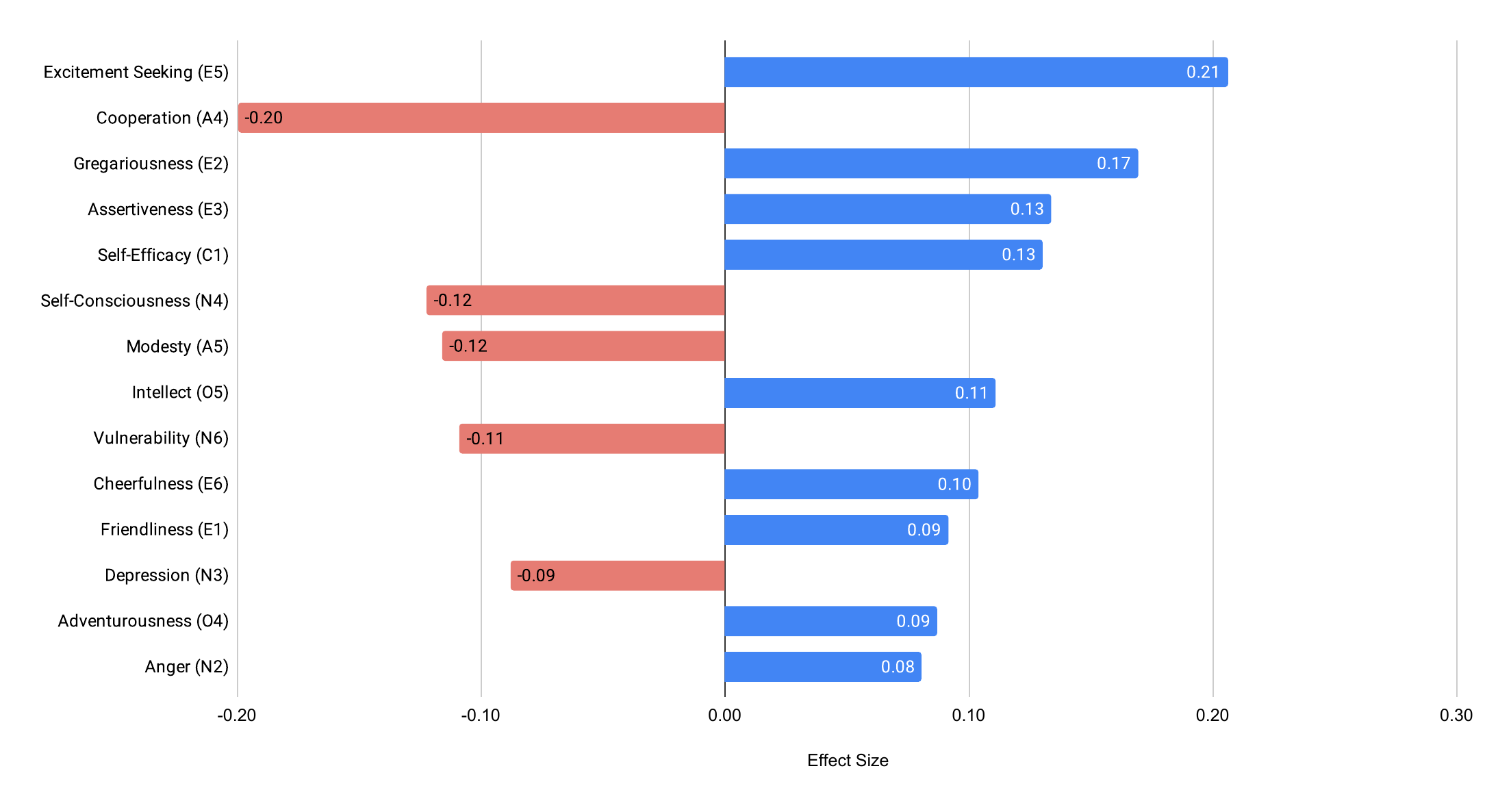} 

	\caption{\textbf{Biggest Intergenerational Changes: Baby Boomers compared to Generation X.} Standardised effect sizes (Cohen’s d) illustrating the largest magnitude trait shifts between Generation X and Baby Boomers. Gen X scores higher than Boomers on the traits shown with blue bars and lower on those shown with pink bars. This historical transition is defined by high external engagement alongside lower baseline cooperation. Baby Boomers exhibited significantly higher Extraversion facets than Gen X, specifically Excitement Seeking (d = 0.21), Gregariousness (d = 0.17), and Assertiveness (d = 0.13). Conversely, Baby Boomers demonstrated markedly lower Cooperation (d = -0.20), Modesty (d = -0.12), and Self-Consciousness (d =-0.12).}
	\label{fig:sup_fig3} 
\end{figure}

\clearpage

\begin{table}[htbp]
\hspace*{-3cm}%
\begin{minipage}{\textwidth}
  \caption{\textbf{Full HAPC Model Coefficients for all 30 Personality Facets (2001-2022)} \textmd{presents the complete set of parameters extracted from the H-APC model across all 30 Big Five personality facets. This table systematically isolates the developmental effects of age (including linear, quadratic, and decadal age scaling) from generational cohort membership (Baby Boomers through Generation Z) and historical period effects (measured in three-year windows from 2001 to 2022). Values in \textbf{bold} indicate $p < .001$; \textbf{underline} indicate $p < .01$; \textit{italics} indicate $p < .05$.}}
  \label{tab:hapc_tab}
  
  \scriptsize
  \setlength{\tabcolsep}{2pt}

  \begin{tabular}{lcccccccccccccc}
    \toprule[1pt]\midrule[0.3pt]
    \multirow{2}{*}[-0.8em]{\centering\textbf{Facets}} &
    \multicolumn{3}{c}{\textbf{Age Effects}} &
    \multicolumn{4}{c}{\textbf{Cohort Effects}} &
    \multicolumn{7}{c}{\textbf{Period Effects}} \\
    \cmidrule(lr){2-4} \cmidrule(lr){5-8} \cmidrule(lr){9-15}
     & \multicolumn{1}{c}{Age} & \multicolumn{1}{c}{Age$^2$} &  \multicolumn{1}{c}{Dec.} & \multicolumn{1}{c}{Boomer} & \multicolumn{1}{c}{Gen X} & \multicolumn{1}{c}{Mill.} & \multicolumn{1}{c}{Gen Z} &  \multicolumn{1}{c}{01-04} & \multicolumn{1}{c}{05-07} & \multicolumn{1}{c}{08-10} & \multicolumn{1}{c}{11-13} & \multicolumn{1}{c}{14-16} & \multicolumn{1}{c}{17-19} & \multicolumn{1}{c}{20-22}\\
    \midrule[0.5pt]
    \textit{Openness} &  &  &  &  &  &  &  &  &  &  &  &  &  &  \\
    Imagination (O1) & \textbf{-0.157} & \textbf{0.001} & \textbf{-1.565} & \textit{0.469} & \textbf{0.936} & \textbf{1.207} & \textbf{0.732} & \textbf{1.552} & \textbf{0.117} & \textbf{-0.315} & \textbf{-0.747} & \textbf{0.236} & \textbf{0.438} & \textbf{-1.280} \\
    Artistic Interests (O2) & \textbf{0.188} & \textbf{-0.002} & \textbf{1.879} & -0.145 & 0.186 & \textbf{0.638} & \textbf{0.849} & \textbf{0.885} & \textbf{0.455} & \textbf{0.221} & \textbf{-0.277} & \textbf{-0.482} & \textbf{0.024} & \textbf{-0.826} \\
    Emotionality (O3) & \textbf{0.177} & \textbf{-0.002} & \textbf{1.774} & 0.125 & \underline{0.495} & \textbf{0.780} & \textbf{0.898} & \textbf{1.061} & \textbf{0.411} & \textbf{-0.086} & \textbf{-0.516} & \textbf{-0.397} & \textbf{0.050} & \textbf{-0.522} \\
    Adventurousness (O4) & \textbf{0.147} & \textbf{-0.001} & \textbf{1.471} & 0.154 & \textbf{0.707} & \textbf{0.842} & 0.234 & \textbf{0.539} & \textbf{0.205} & \textbf{0.238} & \textbf{-0.109} & \textbf{-0.395} & \textbf{-0.093} & \textbf{-0.384} \\
    Intellect (O5) & \textbf{0.236} & \textbf{-0.002} & \textbf{2.364} & 0.235 & \textbf{1.013} & \textbf{0.814} & 0.236 & \textbf{1.093} & \textbf{0.082} & \textbf{-0.195} & \textbf{-0.785} & \textbf{0.182} & \textbf{0.614} & \textbf{-0.990} \\
    Liberalism (O6) & \textbf{0.078} & \textbf{-0.001} & \textbf{0.781} & -0.136 & 0.081 & \underline{0.596} & -0.020 & \textbf{-1.232} & \textbf{-0.881} & \textbf{-1.061} & \textbf{-1.330} & \textbf{0.923} & \textbf{1.240} & \textbf{2.341} \\
    \midrule[0.5pt]
    \textit{Conscientiousness} &  &  &  &  &  &  &  &  &  &  &  &  &  &  \\
    Self-Efficacy (C1) & \textbf{0.405} & \textbf{-0.004} & \textbf{4.054} & 0.076 & \textbf{0.862} & \textbf{1.357} & \textbf{1.085} & \textbf{0.236} & \textbf{0.738} & \textbf{0.768} & \textbf{0.768} & \textbf{-0.567} & \textbf{-0.304} & \textbf{-1.638} \\
    Orderliness (C2) & \textbf{0.532} & \textbf{-0.006} & \textbf{5.323} & \textbf{-1.905} & \textbf{-2.057} & \textbf{-1.801} & \textbf{-0.995} & \textbf{-1.163} & \textbf{-0.057} & \textbf{0.406} & \textbf{1.016} & \textbf{-0.264} & \textbf{0.017} & \textbf{0.043} \\
    Dutifulness (C3) & \textbf{0.398} & \textbf{-0.004} & \textbf{3.977} & \textbf{-0.796} & \textbf{-0.839} & -0.160 & 0.176 & \textbf{-0.318} & \textbf{0.536} & \textbf{0.536} & \textbf{0.767} & \textbf{-0.682} & \textbf{-0.055} & \textbf{-0.783} \\
    Achievement-Striving (C4) & \textbf{0.496} & \textbf{-0.005} & \textbf{4.955} & \textbf{-0.717} & -0.249 & 0.231 & 0.319 & \textbf{-0.099} & \textbf{0.569} & \textbf{0.768} & \textbf{0.998} & \textbf{-0.374} & \textbf{-0.039} & \textbf{-1.822} \\
    Self-Discipline (C5) & \textbf{0.462} & \textbf{-0.004} & \textbf{4.620} & \textbf{-0.707} & \textbf{-0.744} & \textit{-0.477} & \underline{-0.695} & \textbf{-0.315} & \textbf{0.204} & \textbf{0.641} & \textbf{1.287} & \textbf{-0.773} & \textbf{-0.574} & \textbf{-0.469} \\
    Cautiousness (C6) & \textbf{0.513} & \textbf{-0.005} & \textbf{5.128} & \textbf{-1.034} & \textbf{-1.534} & \textbf{-1.029} & \underline{-0.582} & \textbf{-1.764} & \textbf{-0.236} & \textbf{0.116} & \textbf{0.557} & \textbf{-0.138} & \textbf{0.317} & \textbf{1.148} \\
    \midrule[0.5pt]
    \textit{Extraversion} &  &  &  &  &  &  &  &  &  &  &  &  &  &  \\
    Friendliness (E1) & -0.010 & \textbf{0.001} & -0.097 & 0.181 & \textbf{0.903} & \textbf{1.644} & \textbf{1.119} & \textbf{0.356} & \textbf{1.066} & \textbf{1.095} & \textbf{1.225} & \textbf{-1.394} & \textbf{-1.194} & \textbf{-1.1544} \\
    Gregariousness (E2) & \textbf{-0.217} & \textbf{0.002} & \textbf{-2.167} & \textbf{1.010} & \textbf{2.447} & \textbf{3.294} & \textbf{2.369} & \textbf{0.715} & \textbf{1.242} & \textbf{1.159} & \textbf{1.151} & \textbf{-1.536} & \textbf{-1.654} & \textbf{-1.078} \\
    Assertiveness (E3) & \textbf{0.054} & 0.000 & \textbf{0.544} & \underline{0.555} & \textbf{1.530} & \textbf{1.747} & \textbf{1.260} & \textbf{0.563} & \textbf{0.642} & \textbf{0.590} & \textbf{0.503} & \textbf{-0.665} & \textbf{-0.554} & \textbf{-1.078} \\
    Activity Level (E4) & \textbf{0.132} & \textbf{-0.001} & \textbf{1.325} & \textbf{0.627} & \textbf{1.035} & \textbf{0.855} & \textbf{0.803} & \textbf{0.162} & \textbf{0.453} & \textbf{0.472} & \textbf{0.619} & \textbf{-0.584} & \textbf{-0.516} & \textbf{-0.605} \\
    Excitement-Seeking (E5) & \textbf{-0.442} & \textbf{0.004} & \textbf{-4.421} & \textbf{1.999} & \textbf{3.533} & \textbf{3.982} & \textbf{3.085} & \textbf{1.235} & \textbf{0.289} & \textbf{0.297} & \textbf{0.196} & \textbf{-0.036} & \textbf{-0.191} & \textbf{-1.789} \\
    Cheerfulness (E6) & \underline{-0.017} & \textbf{0.000} & \underline{-0.166} & \textit{0.424} & \textbf{1.172} & \textbf{1.999} & \textbf{1.702} & \textbf{0.841} & \textbf{1.148} & \textbf{1.005} & \textbf{0.794} & \textbf{-0.696} & \textbf{-0.603} & \textbf{-2.490} \\
    \midrule[0.5pt]
    \textit{Agreeableness} &  &  &  &  &  &  &  &  &  &  &  &  &  &  \\
    Trust (A1) & \textbf{0.214} & \textbf{-0.001} & \textbf{2.136} & -0.108 & 0.273 & \textbf{1.152} & \textbf{1.061} & \textbf{0.162} & \textbf{0.831} & \textbf{0.579} & \textbf{0.780} & \textbf{-0.676} & \textbf{-0.552} & \textbf{-1.123} \\
    Morality (A2) & \textbf{0.332} & \textbf{-0.003} & \textbf{3.319} & \textbf{-0.672} & \textbf{-1.009} & \textbf{-0.804} & \underline{-0.548} & \textbf{-0.324} & \textbf{0.161} & \textbf{0.081} & \textbf{0.314} & \textbf{-0.752} & \textbf{-0.119} & \textbf{0.638} \\
    Altruism (A3) & \textbf{0.224} & \textbf{-0.002} & \textbf{2.238} & 0.044 & \textit{0.375} & \textbf{1.063} & \textbf{1.151} & \textbf{0.316} & \textbf{0.590} & \textbf{0.450} & \textbf{0.514} & \textbf{-0.771} & \textbf{-0.319} & \textbf{-0.781} \\
    Cooperation (A4) & \textbf{0.375} & \textbf{-0.004} & \textbf{3.747} & \textbf{-1.671} & \textbf{-2.968} & \textbf{-2.646} & \textbf{-2.146} & \textbf{-0.798} & \textbf{0.117} & \textbf{0.243} & \textbf{0.411} & \textbf{-0.386} & \textbf{-0.137} & \textbf{0.550} \\
    Modesty (A5) & \textbf{-0.082} & \textbf{0.001} & \textbf{-0.824} & \underline{0.531} & -0.252 & \textit{-0.466} & 0.022 & \textbf{-0.564} & \textbf{-0.562} & \textbf{-0.384} & \textbf{-0.142} & \textbf{0.300} & \textbf{0.463} & \textbf{0.888} \\
    Sympathy (A6) & \textbf{0.134} & \textbf{-0.001} & \textbf{1.335} & \textbf{0.613} & \underline{0.500} & \textbf{0.918} & \textbf{0.927} & \textbf{-0.211} & \textbf{0.006} & \textbf{-0.283} & \textbf{-0.187} & \textbf{-0.214} & \textbf{0.229} & \textbf{0.661} \\
    \midrule[0.5pt]
    \textit{Neuroticism} &  &  &  &  &  &  &  &  &  &  &  &  &  &  \\
    Anxiety (N1) & \textbf{-0.108} & -0.000 & \textbf{-1.083} & \underline{-0.552} & \textbf{-1.022} & \textbf{-1.602} & \textbf{-0.896} & \textbf{-0.230} & \textbf{-0.566} & \textbf{-0.614} & \textbf{-0.716} & \textbf{0.663} & \textbf{0.656} & \textbf{0.807} \\
    Anger (N2) & \textbf{-0.288} & \textbf{0.003} & \textbf{-2.879} & \textbf{1.011} & \textbf{1.708} & 0.436 & 0.422 & \textbf{0.891} & \textbf{-0.375} & \textbf{-0.630} & \textbf{-0.911} & \textbf{0.529} & \textbf{0.098} & \textbf{0.398} \\
    Depression (N3) & \textbf{-0.193} & \textbf{0.001} & \textbf{-1.930} & -0.075 & \textbf{-0.887} & \textbf{-1.892} & \textbf{-1.146} & \textbf{1.090} & \textbf{-1.059} & \textbf{-1.434} & \textbf{-1.756} & \textbf{1.321} & \textbf{1.241} & \textbf{0.597} \\
    Self-Consciousness (N4) & \textbf{-0.148} & \textbf{0.000} & \textbf{-1.479} & \underline{-0.531} & \textbf{-1.445} & \textbf{-1.967} & \textbf{-1.026} & \textbf{-0.334} & \textbf{-0.673} & \textbf{-0.676} & \textbf{-0.776} & \textbf{0.950} & \textbf{0.889} & \textbf{0.619} \\
    Immoderation (N5) & \textbf{-0.027} & \textbf{-0.001} & \textbf{-0.269} & \textbf{-0.727} & \textbf{-1.090} & \textbf{-1.746} & \textbf{-1.356} & \textbf{1.127} & \textbf{-0.019} & \textbf{-0.404} & \textbf{-0.875} & \textbf{0.518} & \textbf{0.171} & \textbf{-0.518} \\
    Vulnerability (N6) & \textbf{-0.322} & \textbf{0.002} & \textbf{-3.217} & \underline{-0.573} & \textbf{-1.373} & \textbf{-1.780} & \textbf{-1.218} & \textbf{-0.414} & \textbf{-0.611} & \textbf{-0.701} & \textbf{-0.803} & \textbf{0.710} & \textbf{0.583} & \textbf{1.235} \\
  \midrule[0.3pt]\bottomrule[1pt]
\end{tabular}
\end{minipage}
\end{table}

\clearpage

\subsubsection*{Table S2 Commentary: Generational Trends in Entrepreneurial Personality Traits}

To estimate the prevalence of entrepreneurial traits across different generations, we applied a machine learning classifier developed in our previous research \cite{mccarthy2023impact}. This model was originally trained to identify the distinct phenotypic signature of successful startup founders based on Big Five personality facets. We ran this predictor across the full Global Big5 IPIP-NEO dataset to evaluate individuals across all cohorts.

Because the sample sizes for the Greatest Generation (n = 199) and the Silent Generation (n = 2,628) are comparatively small, our analysis primarily focuses on the highly robust samples from the four post-war cohorts.

The predictive modelling reveals a clear and consistent downward trajectory. The proportion of individuals exhibiting the personality traits of successful entrepreneurs peaks with the Baby Boomers at 10.7\% . This relative prevalence steadily declines with each subsequent generation, falling to 9.9\%  for Generation X, 8.4\% for Millennials, and reaching a low of 6.0\%  for Generation Z.

This intergenerational decline in entrepreneurial traits strongly aligns with the broader behavioural shifts discussed earlier in this paper, particularly the rise of safetyism. As younger cohorts adapt to a shifting cultural environment by substituting excitement seeking with an elevated baseline of cautiousness, the psychological profile associated with high risk venture creation appears to be systematically decreasing.

\begin{table}
  \caption{\textbf{Entrepreneurship Prediction across Generations.} \textmd{}}
  \label{tab:entre_pred_tab}
  \centering
  \begin{tabular}{lccc}
    \toprule[1pt]\midrule[0.3pt]
    \centering\textbf{Generations} &
    \textbf{Prediction} &
    \textbf{Counts} &
    \textbf{Percent} \\
    \midrule[0.5pt]
    The Greatest Generation & Entrepreneur & 6 & 3.0\% \\
    The Silent Generation & Entrepreneur & 255 & 9.7\% \\
    Baby Boomers & Entrepreneur & 4,635 & 10.7\% \\
    Generation X & Entrepreneur & 12,432 & 9.9\% \\
    Millennials & Entrepreneur & 29,177 & 8.4\% \\
    Generation Z & Entrepreneur & 15,300 & 6.0\% \\
  \midrule[0.3pt]\bottomrule[1pt]
\end{tabular}
\end{table}

\end{appendices}

\end{document}